# Wave-Packet Treatment of Neutrino Oscillation Based on the Solution to Dirac Equation

Kelin Wang[1], Zexian Cao[2*]

[1]Department of Modern Physics, University of Science and Technology of China, Hefei 230026, China.

[2]Institute of Physics, Chinese Academy of Sciences, P. O. Box 603, Beijing 100190, China.

Flavor oscillation of traveling neutrinos is treated by solving the one-dimensional Dirac equation for massive fermions. The solutions are given in terms of squeezed coherent state as mutual eigenfunctions of parity operator and the corresponding Hamiltonian, both represented in bosonic creation and annihilation operators. It was shown that a mono-energetic state is non-normalizable, and a normalizable Gaussian wave packet, when of pure parity, cannot propagate. A physical state for a traveling neutrino beam would be represented as a normalizable Gaussian wave packet of equally-weighted mixing of two parities, which has the largest energy-dependent velocity. Based on this wave-packet representation, flavor oscillation of traveling neutrinos can be treated in a strict sense. These results allow the accurate interpretation of experimental data for neutrino oscillation, which is critical in judging whether neutrino oscillation violates CP symmetry.

PACS: 14.60.Pq; 42.50.Dv; 11.30.Er

---

[*] Corresponding author. Tel: 86-10-82649136; E-mail: zxcao@iphy.ac.cn

## I. Introduction

Neutrino oscillation arises from the mixture between the flavor and mass eigenstates of neutrinos—an idea first put forward in 1957 and further developed in 1962[1,2]. As a neutrino, generally created as a flavor eigenstate in the weak decay, propagates through space, there is a changing in mixture of mass states since the three mass states advance at slightly different velocity. The correlation between the flavor eigenstates and the mass eigenstates can be described by using three mixing angles $\theta_{12}, \theta_{23}$, and $\theta_{13}$ and a CP phase. The values of $\theta_{12}$ and $\theta_{23}$ have been determined from solar neutrino experiment and atmospheric neutrino experiment, respectively. Recently, by measuring the loss of electron antineutrinos generated at the Daya Bay reactor, the mixing angle $\theta_{13}$ was determined to be $\theta_{13} \sim 9°$, which is surprisingly large[3]. For this reactor-based experiment, the mixing angle $\theta_{13}$ was extracted via the survival probability of the electron antineutrino $\bar{\nu}_e$ at short distances from the reactor, which was given as $P_{sur} \approx 1 - \sin^2(2\theta_{13}) \sin^2(1.267 \Delta m_{13}^2 \cdot L/E)$, where $E$ is the neutrino energy, $L$ is the travel distance, $\Delta m^2$ is the difference of the squared mass eigenvalues[4]. The survival probability relates the decaying of the electron neutrino to $\theta_{13}$ and $\Delta m_{13}^2$. Noting that $\Delta m_{31}^2 = \Delta m_{32}^2 \pm \Delta m_{21}^2$, and $\Delta m_{12}^2$ is relatively very small ($8 \times 10^{-5}$ eV, that's why it needs a long neutrino traveling time to be determined), while $\Delta m_{23}^2$ ($2.4 \times 10^{-3}$ eV) and $\Delta m_{13}^2$ are larger[5], thus if the measurement time for electron neutrino loss is shorter, the possible effect of $\theta_{12}, \Delta m_{12}^2, \theta_{23}$, and $\Delta m_{23}^2$ can be omitted, and $\Delta m_{13}^2$ takes the value of $\Delta m_{23}^2$ without further clarification. As the measurement at Daya Bay was done at a closer distance—reactor-detector distance is a little less than 2 km for a largest neutrino loss, this is provably a good approximation.

The accurate determination of the three mixing angles is of essential importance to the understanding of neutrino flavor oscillation, which in turn helps understand the possible CP symmetry breaking. Now it has been established that we do have three nonzero mixing angles. More measurements at a few places are planned to improve the accuracy of $\theta_{13}$ as it bears some significance to both particle physics and cosmology. The accuracy of mixing angles from oscillation measurement hangs on the availability of a more accurate expression of survival probability, of which the implementation needs fewer assumptions, thus allowing a flexible experimental arrangement, e.g., the source-detector distance is no more a critical prerequisite. Such an expression for survival probability should be obtained in a strict sense.

However, we noticed that the surviving probability was given with regard to a single energy value for the neutrino, but the true neutrino beam may have a wide energy distribution. For example, the energy spectrum of the electron antineutrinos from the reactor at Daya Bay peaks at ~4.0 MeV with a half width of comparable value[6]. This fact perhaps will largely limit the applicability if not invalidate the formula, at least it brings some error to the calculation especially when also some other parameters are unknown. Furthermore, it is worth noting that the quantum state of a mono-energetic fermion cannot be normalized; and in quantum mechanics wavefunctions for the interaction-free beam of identical particles such as neutrino is normalized to particle number, whereas the wave function used in calculating the survival probability is that for a single particle. Some new insights might be yielded when the problem is reconsidered from scratch, i.e., starting from the Dirac equation.

The true physical state for a neutrino beam should be a wave packet with the energy distributed in a finite range. Thus in order to calculate the mixing state of neutrino beam at an arbitrary distance from the source and to compare with measurement, the question should be discussed with the states of neutrinos properly represented in wave packet. This is the motivation for the current article. In fact, wave packet approach to the neutrino oscillation has been proposed with different motivations[7-10], e.g., to cope with the uncertainty principle, but it has not got the deserved attention. In the current paper the following points will be discussed. First the wave packet for neutrinos will be constructed on the solution to one-dimensional Dirac equation for massive neutrino, or on the basis of mutual eigenfunctions of parity operator and the corresponding Hamiltonian, to be more specific. Then the propagation of particles in such a wave packet state will be discussed and the $P_{sur}$ for all the neutrino flavors are to be obtained, which is expected to help improve the data interpretation. The difference in the surviving probabilities for neutrino and antineutrino, which can be accessible only when a precise data interpretation is done, is critical to the investigation of possible CP violation.

**II. Formalism and discussion**

Neutrinos generated in a source usually propagate in all directions. But, for a detector placed a few kilometers away from the source, the solid angle subtended by the detector is very small that in principle the neutrinos arriving at the detector have a vanishing transverse momentum, i.e., the propagation of neutrinos can be approximately treated as a one-dimensional problem.

As is well-known that out of the three flavors of neutrios at least two are massive, therefore we start with the Dirac equation for massive fermions, where the state of the neutrinos is represented in the 4-component form (by setting $c=1$, and $\hbar=1$)

$$(\alpha_3 \cdot \hat{p}_z + m\beta) \begin{pmatrix} |\Phi_1\rangle \\ |\Phi_2\rangle \\ |\Phi_3\rangle \\ |\Phi_4\rangle \end{pmatrix} = E \begin{pmatrix} |\Phi_1\rangle \\ |\Phi_2\rangle \\ |\Phi_3\rangle \\ |\Phi_4\rangle \end{pmatrix} \qquad (1).$$

With the explicit expression for $\alpha_3$ and $\beta$, this equation can be further rewritten as

$$\begin{aligned} \hat{p}|\Phi_3\rangle + m|\Phi_1\rangle &= E|\Phi_1\rangle \\ -\hat{p}|\Phi_4\rangle + m|\Phi_2\rangle &= E|\Phi_2\rangle \\ \hat{p}|\Phi_1\rangle - m|\Phi_3\rangle &= E|\Phi_3\rangle \\ -\hat{p}|\Phi_2\rangle - m|\Phi_4\rangle &= E|\Phi_4\rangle \end{aligned} \qquad (2).$$

We see that $(\Phi_1, \Phi_3)$ are decoupled from $(\Phi_2, \Phi_4)$, with the former referred to the spin-up state and the latter to the spin-down state. As problems here concerned are not related to spin, so only the $(\Phi_1, \Phi_3)$ components will be discussed below. For clarity, by replacing $|\Phi_1\rangle$ and $|\Phi_3\rangle$ with $|\varphi_1\rangle$ and $|\varphi_2\rangle$, respectively, we arrive at the following two equations

$$\begin{aligned} \hat{p}|\varphi_2\rangle + m|\varphi_1\rangle &= E|\varphi_1\rangle \\ \hat{p}|\varphi_1\rangle - m|\varphi_2\rangle &= E|\varphi_2\rangle \end{aligned} \qquad (3).$$

Now, let's introduce the bosonic creation and annihilation operators to represent the momentum and the position,

$$\hat{p} = \frac{1}{\sqrt{2}\Delta}(b + b^+)$$
$$\hat{z} = \frac{i\Delta}{\sqrt{2}}(b - b^+) \qquad (4)$$

Then the equations (3) can be rewritten as

$$\sigma_x \frac{1}{\sqrt{2}\Delta}(b+b^+)\begin{pmatrix}|\varphi_1\rangle\\|\varphi_2\rangle\end{pmatrix} + m\sigma_z \begin{pmatrix}|\varphi_1\rangle\\|\varphi_2\rangle\end{pmatrix} = E\begin{pmatrix}|\varphi_1\rangle\\|\varphi_2\rangle\end{pmatrix} \qquad (5)$$

Or it can be said that we are dealing with the eigenvalue problem for the Hamiltonian

$$\hat{H} = A\sigma_x(b+b^+) + m\sigma_z, \text{ where } A = \frac{1}{\sqrt{2}\Delta} \text{ is a parameter in dimension of momentum,}$$

and $\sigma_x$ and $\sigma_z$ are Pauli matrices.

## A. Mutual eigenstates for parity operator and Hamiltonian

For this system we now introduce the parity operator

$$\hat{\Pi} = \exp[i\pi(\sigma_z + \tfrac{1}{2} + b^+b)] \qquad (6)$$

which is a conservative quantity as it has $[\hat{\Pi}, \hat{H}] = 0$. The two eigenstates for the parity operator can be represented in the form of squeezed coherent state as follows

$$|\Pi^+\rangle = \begin{pmatrix} (e^{-\frac{1}{2}b^+b^+ + \alpha b^+} - e^{-\frac{1}{2}b^+b^+ - \alpha b^+})|0\rangle \\ g^+(e^{-\frac{1}{2}b^+b^+ + \alpha b^+} + e^{-\frac{1}{2}b^+b^+ - \alpha b^+})|0\rangle \end{pmatrix} \qquad (7a)$$

$$|\Pi^-\rangle = \begin{pmatrix} (e^{-\frac{1}{2}b^+b^+ + \alpha b^+} + e^{-\frac{1}{2}b^+b^+ - \alpha b^+})|0\rangle \\ g^-(e^{-\frac{1}{2}b^+b^+ + \alpha b^+} - e^{-\frac{1}{2}b^+b^+ - \alpha b^+})|0\rangle \end{pmatrix} \qquad (7b)$$

where $g^\pm$ are parameters to be determined. To prove this, we see that

$$(e^{-\frac{1}{2}b^+b^+ +\alpha b^+} - e^{-\frac{1}{2}b^+b^+ -\alpha b^+})|0\rangle = e^{-\frac{1}{2}b^+b^+}(e^{\alpha b^+} - e^{-\alpha b^+})|0\rangle$$

$$= e^{-\frac{1}{2}b^+b^+}(\sum_n \frac{2(\alpha)^{2n+1}}{(2n+1)!}(b^+)^{2n+1})|0\rangle = \sum_m f_{2m+1}(b^+)^{2m+1}|0\rangle$$

containing only odd-number particle states, and

$$(e^{-\frac{1}{2}b^+b^+ +\alpha b^+} + e^{-\frac{1}{2}b^+b^+ -\alpha b^+})|0\rangle = e^{-\frac{1}{2}b^+b^+}(\sum_n \frac{2(\alpha)^{2n}}{(2n)!}(b^+)^{2n})|0\rangle$$

$$= \sum_m \varphi_{2m}(b^+)^{2m}|0\rangle$$

containing only even-number particle states, thus it has $\hat{\Pi}|\Pi^+\rangle = |\Pi^+\rangle$, i.e., the parity of the state $|\Pi^+\rangle$ is even. Similarly, it can be proved that the parity of the state $|\Pi^-\rangle$ is odd. Since the parity operator has eigenvalues of $\pm 1$, the eigenstates of the Hamiltonian can be grouped into two branches according to the two eigenvalues of the parity operator.

As $[\hat{\Pi}, \hat{H}] = 0$, the states in (7) can be also the eigenstates of the Hamiltonian if the parameter $\alpha$ and coefficients $g^\pm$ are properly chosen. To make the state $|\Pi^+\rangle$ in eq.(7a) a stationary solution of the Hamiltonian,

$$[A\begin{pmatrix} 0 & 1 \\ 1 & 0 \end{pmatrix}(b+b^+) + m\begin{pmatrix} 1 & 0 \\ 0 & -1 \end{pmatrix}]\begin{pmatrix} (e^{-\frac{1}{2}b^+b^+ +\alpha b^+} - e^{-\frac{1}{2}b^+b^+ -\alpha b^+})|0\rangle \\ g^+(e^{-\frac{1}{2}b^+b^+ +\alpha b^+} + e^{-\frac{1}{2}b^+b^+ -\alpha b^+})|0\rangle \end{pmatrix}$$

$$= E\begin{pmatrix} (e^{-\frac{1}{2}b^+b^+ +\alpha b^+} - e^{-\frac{1}{2}b^+b^+ -\alpha b^+})|0\rangle \\ g^+(e^{-\frac{1}{2}b^+b^+ +\alpha b^+} + e^{-\frac{1}{2}b^+b^+ -\alpha b^+})|0\rangle \end{pmatrix} \quad (8)$$

One sees that it requires $A\alpha g^+ = (E-m)$ from the upper component, and from the low component $A\alpha = (E+m)g^+$, thus it has $E(\alpha)^2 = m^2 + A^2\alpha^2$ or $E(\alpha) = \pm\sqrt{m^2 + A^2\alpha^2}$. Furthermore, for the positive energy solution, $E(\alpha) = \sqrt{m^2 + A^2\alpha^2}$, it has $g^+ = \dfrac{A\alpha}{\sqrt{m^2 + A^2\alpha^2} + m}$; while for the negative energy solution, $E(\alpha) = -\sqrt{m^2 + A^2\alpha^2}$, it has $g^+ = -\dfrac{\sqrt{m^2 + A^2\alpha^2} + m}{A\alpha}$.

For the negative parity, the same results can be obtained. Or in other words, the energy eigenstates are doubly degenerated.

**B. Wave packet**

We noticed that the mutual eigenstates of the parity operator and the Hamiltonian are non-normalizable. The true physical state should be a wave packet comprising a group of such states of variable $\alpha$. It can be easily proven that the Gaussian wave packets given below

$$\left|\Phi^+\right\rangle = \int d\alpha\, e^{-\gamma(\alpha-\alpha_0)^2} \begin{pmatrix} (e^{-\frac{1}{2}b^+b^+ + \alpha b^+} - e^{-\frac{1}{2}b^+b^+ - \alpha b^+})|0\rangle \\ g^+(e^{-\frac{1}{2}b^+b^+ + \alpha b^+} + e^{-\frac{1}{2}b^+b^+ - \alpha b^+})|0\rangle \end{pmatrix} \quad (9a)$$

$$\left|\Phi^-\right\rangle = \int d\alpha\, e^{-\gamma(\alpha-\alpha_0)^2} \begin{pmatrix} (e^{-\frac{1}{2}b^+b^+ + \alpha b^+} + e^{-\frac{1}{2}b^+b^+ - \alpha b^+})|0\rangle \\ g^-(e^{-\frac{1}{2}b^+b^+ + \alpha b^+} - e^{-\frac{1}{2}b^+b^+ - \alpha b^+})|0\rangle \end{pmatrix} \quad (9b)$$

for example, preserve the parity. In the following, the discussion is limited to the positive energy solution, for which $g^+ = g^- = \dfrac{A\alpha}{\sqrt{m^2 + A^2\alpha^2} + m}$ in (9), thus denoted since now on simply as $g(m)$.

However, it needs be pointed out that neutrinos cannot propagate with this kind of wave packet of pure parity, since, supposing that at $t = 0$ the neutrino is found in a state of $\left|\Phi^+\right\rangle$, then at any following moment $t$ the state will be

$$\left|\Phi^+(t)\right\rangle = \int d\alpha\, e^{-\gamma(\alpha-\alpha_0)^2} e^{-iE(\alpha)t} \begin{pmatrix} (e^{-\frac{1}{2}b^+b^+ + \alpha b^+} - e^{-\frac{1}{2}b^+b^+ - \alpha b^+})|0\rangle \\ g(m)(e^{-\frac{1}{2}b^+b^+ + \alpha b^+} + e^{-\frac{1}{2}b^+b^+ - \alpha b^+})|0\rangle \end{pmatrix} \quad (10),$$

but it has $\frac{1}{\langle \Phi^+(t)|\Phi^+(t)\rangle}\langle \Phi^+(t)|\hat{z}|\Phi^+(t)\rangle = 0$. This is to say that in such a wave packet state of pure parity the propagation velocity of particle is zero. Only those states of mixed parity can propagate. Similar conclusion is also drawn for the Zitterbewegung of electrons, to which the Hamiltonian and the parity operator differ from those concerned here only by sign at one place[11].

A wave packet state for neutrinos that propagate thus must be of mixed parities. For the wave packet with equal-weight mixed parities

$$|\Phi\rangle = \tfrac{1}{2}(|\Phi^+\rangle + |\Phi^-\rangle) = \int d\alpha e^{-\gamma(\alpha-\alpha_0)^2} \begin{pmatrix} e^{-\tfrac{1}{2}b^+b^+ + \alpha b^+}|0\rangle \\ g(m) e^{-\tfrac{1}{2}b^+b^+ + \alpha b^+}|0\rangle \end{pmatrix} \quad (11),$$

it has the maximum velocity of $\sqrt{1 - m^2/E^2} \sim 1$, as can be expected.

**C. Flavor oscillation**

According to the standard theory, the neutrinos of a given flavor are in the mixed mass eigenstates. Accordingly, here we represent the initial state of a neutrino of a specific flavor with a Gaussian wave packet which has been constructed on the mixed mass eigenstates. Taking $\nu_e$ for example, the initial state is

$$|t=0\rangle = \int d\alpha e^{-\gamma(\alpha-\alpha_0)^2} [\cos\theta_{13}\cos\theta_{12}|\nu_1(\alpha)\rangle$$
$$+ \cos\theta_{13}\sin\theta_{12}e^{i\Phi_{12}}|\nu_2(\alpha)\rangle \quad (12),$$
$$+ \sin\theta_{13}e^{i\Phi_{13}}|\nu_3(\alpha)\rangle]$$

where $|\nu_i(\alpha)\rangle = \begin{pmatrix} e^{-\tfrac{1}{2}b_i^+ b_i^+ + \alpha b_i^+}|0\rangle \\ g(m_i) e^{-\tfrac{1}{2}b_i^+ b_i^+ + \alpha b_i^+}|0\rangle \end{pmatrix}, i=1,2,3$, denotes the three energy eigenstates for neutrino. To emphasize, all these energy eigenstates for neutrino adopt the form of

wave packet of equal-weighted parities in eq.(11).

The state in (12) is not yet normalized. Since the rest mass of the neutrinos is very small in comparison to the kinetic energy $A\alpha_0$, clearly it has $m_i \ll A\alpha$, thus we can adopt the approximation $g_i = \frac{A\alpha}{\sqrt{m_i^2 + A^2\alpha^2} + m_i} \approx 1 - \frac{m_i}{A\alpha}$. Furthermore, only those $\alpha$ in the neighborhood of $\alpha_0$ contribute some to the integral in (12), consequently we can assume $g_i \approx 1 - \frac{m_i}{A\alpha_0}$ in calculating the integral. Considering the fact that the operators $b_1$ ($b_1^+$), $b_2$ ($b_2^+$), and $b_3$ ($b_3^+$) are mutually independent, thus the normalization factor was to be approximated by

$$N = \langle t=0|t=0\rangle = [\cos^2\theta_{13}\cos^2\theta_{12}(1+g_1^2) + \cos^2\theta_{13}\sin^2\theta_{12}(1+g_2^2) \\ + \sin^2\theta_{13}(1+g_3^2)] \cdot \sqrt{\frac{4}{4\gamma-1}}\pi e^{\frac{2\gamma}{4\gamma-1}\alpha_0^2} \qquad (13)$$

When a flavor neutrino propagates, its state evolves that to each mass eigenstate in (12) a phase factor $e^{-iE_i(\alpha)t}$, where $E_i(\alpha) = \sqrt{m_i^2 + A^2\alpha^2} \approx A\alpha + \frac{m_i^2}{2A\alpha}$, is added. Consequently, at any later moment $t$ the state $|t\rangle$, by using the mixing matrix $U$ [12] (see appendix), is given as

$$|t\rangle = N^{-1/2}\int d\alpha e^{-\gamma(\alpha-\alpha_0)^2}[\cos\theta_{13}\cos\theta_{12}e^{-iE_1t}|v_1(\alpha)\rangle \\ + \cos\theta_{13}\sin\theta_{12}e^{i\Phi_{12}}e^{-iE_2t}|v_2(\alpha)\rangle \\ + \sin\theta_{13}e^{i\Phi_{13}}e^{-iE_3t}|v_3(\alpha)\rangle] \qquad (14)$$

By using the transformation for $v_i \to v_{(e,\mu,\tau)}$, i.e., the inverse mixing matrix, the state $|t\rangle$ which, say, denotes a state for $v_e$ can then be expanded in the states of all three flavors

$$|t\rangle = N^{-1/2} \int d\alpha e^{-\gamma(\alpha-\alpha_0)^2} [\cos\theta_{13}\cos\theta_{12}e^{-iE_1 t}\{\cos\theta_{12}\cos\theta_{13}|\nu_e(\alpha)\rangle$$
$$-(e^{-i(\Phi_{13}-\Phi_{23})}\cos\theta_{12}\sin\theta_{13}\sin\theta_{23}+e^{-i\Phi_{12}}\cos\theta_{23}\sin\theta_{12})|\nu_\mu(\alpha)\rangle$$
$$+(e^{-i(\Phi_{12}+\Phi_{23})}\sin\theta_{12}\sin\theta_{23}-e^{-i\Phi_{13}}\cos\theta_{23}\cos\theta_{12}\sin\theta_{13})|\nu_\tau(\alpha)\rangle\}$$
$$+\cos\theta_{13}\sin\theta_{12}e^{i\Phi_{12}}e^{-iE_2 t}\{e^{-i\Phi_{12}}\cos\theta_{13}\sin\theta_{12}|\nu_e(\alpha)\rangle$$
$$+(\cos\theta_{12}\cos\theta_{23}-e^{-i(\Phi_{12}+\Phi_{23}-\Phi_{13})}\sin\theta_{12}\sin\theta_{13}\sin\theta_{23})|\nu_\mu(\alpha)\rangle$$
$$-(e^{i\Phi_{23}}\cos\theta_{12}\sin\theta_{23}+e^{-i(\Phi_{12}-\Phi_{13})}\cos\theta_{23}\sin\theta_{12}\sin\theta_{13})|\nu_\tau(\alpha)\rangle\}$$
$$+\sin\theta_{13}e^{i\Phi_{13}}e^{-iE_3 t}\{e^{-i\Phi_{13}}\sin\theta_{13}|\nu_e(\alpha)\rangle+e^{-i\Phi_{23}}\cos\theta_{12}\sin\theta_{23}|\nu_\mu(\alpha)\rangle$$
$$+\cos\theta_{13}\cos\theta_{23}|\nu_\tau(\alpha)\rangle\}$$
(15)

Thus for a $\nu_e$, the probability amplitudes for it to be in the state of $|\nu_e(\alpha)\rangle$, $|\nu_\mu(\alpha)\rangle$ and $|\nu_\tau(\alpha)\rangle$, at a later moment $t$, are

$$\psi_e(\alpha,t) = N^{-1/2}e^{-\gamma(\alpha-\alpha_0)^2}[\cos^2\theta_{13}\cos^2\theta_{12}e^{-iE_1 t}$$
$$+\cos^2\theta_{13}\sin^2\theta_{12}e^{-iE_2 t}+\sin^2\theta_{13}e^{-iE_3 t}]$$
(16a),

$$\psi_\mu(\alpha,t) = N^{-1/2}e^{-\gamma(\alpha-\alpha_0)^2}[-\cos^2\theta_{12}\cos\theta_{13}\sin\theta_{13}\sin\theta_{23}e^{i(\Phi_{13}-\Phi_{23})}$$
$$-\cos\theta_{12}\cos\theta_{13}\cos\theta_{23}\sin\theta_{12}e^{i\Phi_{12}})e^{-iE_1 t}+(\cos\theta_{13}\cos\theta_{23}\cos\theta_{12}\sin\theta_{12}e^{i\Phi_{12}}$$
$$-\cos\theta_{13}\sin^2\theta_{12}\sin\theta_{13}\sin\theta_{23}e^{-i(\Phi_{23}-\Phi_{13})})e^{-iE_2 t}+\cos\theta_{13}\sin\theta_{23}\sin\theta_{13}e^{-i(\Phi_{23}-\Phi_{13})}e^{-iE_3 t}]$$
(16b),

$$\psi_\tau(\alpha,t) = N^{-1/2}e^{-\gamma(\alpha-\alpha_0)^2}[(-\cos^2\theta_{12}\cos\theta_{13}\cos\theta_{23}\sin\theta_{13}e^{i\Phi_{13}}$$
$$+\cos\theta_{12}\cos\theta_{13}\sin\theta_{23}\sin\theta_{12}e^{i(\Phi_{12}+\Phi_{23})})e^{-iE_1 t}-(\cos\theta_{13}\sin\theta_{23}\cos\theta_{12}\sin\theta_{12}e^{i(\Phi_{12}+\Phi_{23})}$$
$$+\cos\theta_{13}\sin^2\theta_{12}\sin\theta_{13}\cos\theta_{23}e^{i\Phi_{13}})e^{-iE_2 t}+\cos\theta_{13}\cos\theta_{23}\sin\theta_{13}e^{i\Phi_{13}}e^{-iE_3 t}]$$
(16c),

respectively. From these amplitudes a strict express for survival probability for the electron neutrino can be obtained.

Two main conclusions can be drawn from the above argument. First, since the energy is uniquely determined by the parameter $\alpha$, one needs just calculate, for example, $\rho_e(\alpha,t)$ to get $\rho_e(E(\alpha),t)$. At any moment $t$, the probability density is

$$\rho_e(\alpha,t) = N^{-1} e^{-2\gamma(\alpha-\alpha_0)^2} [\cos^4\theta_{12}\cos^4\theta_{13} + \cos^4\theta_{13}\sin^4\theta_{12} + \sin^4\theta_{13}$$
$$+ \tfrac{1}{2}\cos^2\theta_{13}\sin^2(2\theta_{12}) \cdot \cos(\frac{\Delta m_{12}^2}{2A\alpha}t) \qquad (17).$$
$$+ \tfrac{1}{2}\cos^2\theta_{12}\sin^3(2\theta_{13})\cos(\frac{\Delta m_{13}^2}{2A\alpha}t) + \tfrac{1}{2}\sin^2\theta_{12}\sin^2(2\theta_{13})\cos(\frac{\Delta m_{23}^2}{2A\alpha}t)$$

As to the flavor oscillation, let's calculate, say, the probability density for $\nu_e \to \nu_\mu$, one just needs calculate $\rho_\mu(\alpha,t)$ from eq.(16b):

$$\rho_\mu(\alpha,t) = N^{-1} e^{-2\gamma(\alpha-\alpha_0)^2}[\cos^2\theta_{12}\cos^2\theta_{13}\cos^2\theta_{23}\sin^2\theta_{12} + \cos^4\theta_{12}\cos^2\theta_{13}\sin^2\theta_{13}\sin^2\theta_{23}$$
$$+2\cos^3\theta_{12}\cos^2\theta_{13}\cos\theta_{23}\sin\theta_{12}\sin\theta_{13}\sin\theta_{23}\cos(\Phi_{12}+\Phi_{23}-\Phi_{13})$$
$$+\cos^2\theta_{12}\cos^2\theta_{13}\cos^2\theta_{23}\sin^2 2\theta_{12} + \sin^4\theta_{12}\cos^2\theta_{13}\sin^2\theta_{13}\sin^2\theta_{23}$$
$$-2\cos^2\theta_{13}\cos\theta_{23}\cos\theta_{12}\sin^3\theta_{12}\sin\theta_{13}\sin\theta_{23}\cos(\Phi_{12}+\Phi_{23}-\Phi_{13})$$
$$+\cos^2\theta_{13}\sin^2\theta_{13}\sin^2\theta_{23}$$
$$-2\cos^2\theta_{12}\cos^2\theta_{13}\cos^2\theta_{23}\sin^2\theta_{12}\cos(\frac{\Delta m_{12}^2}{2A\alpha}t)$$
$$+2\cos\theta_{12}\cos^2\theta_{13}\cos\theta_{23}\sin^3\theta_{12}\sin\theta_{13}\sin\theta_{23}\cos(\frac{\Delta m_{12}^2}{2A\alpha}t - \Phi_{12}-\Phi_{23}+\Phi_{13})$$
$$-2\cos^2\theta_{13}\cos^2\theta_{12}\sin^2\theta_{12}\sin^2\theta_{13}\sin^2\theta_{23}\cos(\frac{\Delta m_{12}^2}{2A\alpha}t)$$
$$-2\cos^2\theta_{12}\cos^2\theta_{13}\cos\theta_{23}\sin\theta_{12}\sin\theta_{13}\sin\theta_{23}\cos(\frac{\Delta m_{13}^2}{2A\alpha}t - \Phi_{12}-\Phi_{23}+\Phi_{13})$$
$$+2\cos^2\theta_{12}\cos^2\theta_{13}\sin^2\theta_{13}\sin^2\theta_{23}\cos(\frac{\Delta m_{13}^2 t}{2A\alpha} - \Phi_{12}-\Phi_{13}+\Phi_{23})$$
$$+2\cos\theta_{12}\cos^2\theta_{13}\cos\theta_{23}\sin\theta_{12}\sin\theta_{13}\sin\theta_{23}\cos(\frac{\Delta m_{23}^2}{2A\alpha}t - \Phi_{12}-\Phi_{13}+\Phi_{23}) \qquad (18).$$
$$-2\cos^2\theta_{13}\sin^2\theta_{12}\sin^2\theta_{13}\sin^2\theta_{23}\cos(\frac{\Delta m_{23}^2}{2A\alpha}t)]$$

The discussion above is given in $\alpha$, but just remember that $E_i = \sqrt{m_i^2 + A^2\alpha^2}$, then the $\alpha$ can be replaced with E. As a rough approximation $E \approx A\alpha$. For practical implementation, the width of wave packet $\gamma$ and the parameter $\alpha_0$ can be input as experimental parameters.

Integrating eq.(17) over $\alpha$, we can obtain the surviving probability of $\nu_e$ at any later moment $t$. The formula (18) gives the probability for the $\nu_e \to \nu_\mu$. To get the probability for $\bar{\nu}_e \to \bar{\nu}_\mu$, one just replaces the parameters ($\Phi_{12}, \Phi_{23}, \Phi_{13}$) with ($-\Phi_{12}, -\Phi_{23}, -\Phi_{13}$), thus

$$\rho_\mu(\alpha,t) - \rho_{\bar{\mu}}(\alpha,t) = 4N^{-1} e^{-2\gamma(\alpha-\alpha_0)^2} [\cos\theta_{12}\cos^2\theta_{13}\cos\theta_{23}\sin^3\theta_{12}\sin\theta_{13}\sin\theta_{23}\sin(\frac{\Delta m_{12}^2}{2A\alpha}t)$$
$$+\cos^2\theta_{12}\cos^2\theta_{13}\sin\theta_{13}\sin\theta_{23}(\sin\theta_{13}\sin\theta_{23}-\cos\theta_{23}\sin\theta_{12})\sin(\frac{\Delta m_{13}^2}{2A\alpha}t) \quad (19).$$
$$+\cos\theta_{12}\cos^2\theta_{13}\cos\theta_{23}\sin\theta_{12}\sin\theta_{13}\sin\theta_{23}\sin(\frac{\Delta m_{23}^2}{2A\alpha}t)]\sin(\Phi_{12}+\Phi_{23}-\Phi_{13})$$

If there is CP violation for the neutrino oscillation, this result thus can be used to decode the invariant combination of CP phases $\Phi_{12}+\Phi_{23}-\Phi_{13}$ [13] from experimental data.

### III. Summary

In summary, neutrino oscillation was treated by solving the one-dimensional Dirac equation for massive fermions. The solution can be represented in terms of squeezed coherent state as the mutual eigenfunction of the parity operator and the corresponding hamiltonian, but it is non-normalizable. The physical state has to be in the form of wave packet. Furthermore, no state of pure parity can propagate while that of equally-weighted mixing of two parities has the largest energy-dependent velocity $v = A\alpha/m$. The flavor oscillation of a traveling neutrino beam can be obtained in a strict sense. These results can help treat the experimental data for neutrino flavor oscillation for a more accurate determination of the mixing angles, which in turn can be more confirmative about the CP violation.

**Acknowledgments**

One of the authors, Cao Zexian, thanks the financial support from the Natural Science Foundation of China grant nos. 51172272 and 10974227, and from the Chinese Academy of Sciences.

**Appendix**

The mass eigenstates and energy eigenstates is related via the mixing matrix U

$$\begin{pmatrix} \nu_e \\ \nu_\mu \\ \nu_\tau \end{pmatrix} = U \begin{pmatrix} \nu_1 \\ \nu_2 \\ \nu_3 \end{pmatrix} \quad (A1)$$

Where 1, 2, 3 denote neutrino states with a definite mass.

For the case of three neutrinos, following Ref.[7], the lepton mixing matrix can be parametrized as $U = \omega_{23}\omega_{13}\omega_{12}$, where each factor $\omega$ is effectively $2 \times 2$, characterized by an angle and a CP phase, e.g.,

$$\omega_{13} = \begin{pmatrix} \cos\theta_{13} & 0 & e^{i\Phi_{13}}\sin\theta_{13} \\ 0 & 1 & 0 \\ -e^{-i\Phi_{13}}\sin\theta_{13} & 0 & \cos\theta_{13} \end{pmatrix} \quad (A2).$$

Such a symmetrical parameterization of the lepton mixing matrix, can be written as

$$U = \begin{pmatrix} \cos\theta_{12}\cos\theta_{13} & e^{i\Phi_{12}}\sin\theta_{12}\sin\theta_{13} & e^{i\Phi_{13}}\sin\theta_{13} \\ -e^{-i\Phi_{12}}\sin\theta_{12}\cos\theta_{23} - e^{i(\Phi_{23}-\Phi_{13})}\cos\theta_{12}\sin\theta_{13}\sin\theta_{23} & \cos\theta_{12}\cos\theta_{23} - e^{i(\Phi_{12}+\Phi_{23}-\Phi_{13})}\sin\theta_{12}\sin\theta_{13}\sin\theta_{23} & e^{i\Phi_{23}}\cos\theta_{13}\sin\theta_{23} \\ e^{-i(\Phi_{12}+\Phi_{23})}\sin\theta_{12}\sin\theta_{23} & -e^{-i\Phi_{23}}\cos\theta_{12}\sin\theta_{23} - e^{i(\Phi_{12}-\Phi_{13})}\sin\theta_{12}\sin\theta_{13}\cos\theta_{23} & \cos\theta_{13}\cos\theta_{23} \end{pmatrix}$$

$$(A3)$$

Where the invariant combination $\delta = \Phi_{12} + \Phi_{23} - \Phi_{13}$ corresponds to the Dirac phase.

The inversed matrix reads

$$U^{-1} = \begin{pmatrix} \cos\theta_{12}\cos\theta_{13} & -e^{i(\Phi_{13}-\Phi_{23})}\cos\theta_{12}\sin\theta_{13}\sin\theta_{23} - e^{i\Phi_{12}}\sin\theta_{12}\cos\theta_{23} & e^{i(\Phi_{12}+\Phi_{23})}\sin\theta_{12}\sin\theta_{23} - e^{i\Phi_{13}}\cos\theta_{23}\cos\theta_{12}\sin\theta_{13} \\ e^{-i\Phi_{12}}\sin\theta_{12}\cos\theta_{13} & \cos\theta_{12}\cos\theta_{23} - e^{-i(\Phi_{12}+\Phi_{23}-\Phi_{13})}\sin\theta_{12}\sin\theta_{13}\sin\theta_{23} & -e^{i\Phi_{23}}\cos\theta_{12}\sin\theta_{23} - e^{-i(\Phi_{12}-\Phi_{13})}\cos\theta_{23}\sin\theta_{12}\sin\theta_{13} \\ e^{-i\Phi_{13}}\sin\theta_{13} & e^{-i\Phi_{23}}\cos\theta_{13}\sin\theta_{23} & \cos\theta_{13}\cos\theta_{23} \end{pmatrix}$$

$$(A4)$$